\documentclass{article}

\usepackage{amsfonts}

\begin{document}


\def\Bid{{\mathchoice {\rm {1\mskip-4.5mu l}} {\rm
{1\mskip-4.5mu l}} {\rm {1\mskip-3.8mu l}} {\rm {1\mskip-4.3mu l}}}}


\newcommand{\beq}{\begin{equation}}
\newcommand{\eeq}{\end{equation}}
\newcommand{\bea}{\begin{eqnarray}}
\newcommand{\eea}{\end{eqnarray}}
\newcommand{\eL}{{\cal L}}
\newcommand{\J}{\bf J}
\newcommand{\bP}{\bf P}
\newcommand{\G}{\bf G}
\newcommand{\K}{\bf K}
\newcommand{\M}{{\cal M}}
\newcommand{\bu}{\bf u}
\newcommand{\la}{\lambda}
\newcommand{\half}{\frac{1}{2}}

\begin{flushright}
DOE-ER-40757-104\\
UTEXAS-HEP-97-18
\end{flushright}

\hfil\hfil

\begin{center}
{\Large {\bf Differential Geometry on SU(3) with Applications to Three State Systems}}
\end{center}

\hfil\break

\begin{center}
{\bf by Mark Byrd \footnote{mbyrd@physics.utexas.edu} \\}
{\it Center for Particle Physics \\
University of Texas at Austin \\
Austin, Texas 78712-1081}
\end{center}

\begin{abstract}
The left and right invariant vector fields are calculated in an ``Euler angle'' type parameterization for the group manifold of $SU(3)$, referred to here as Euler coordinates.  The corresponding left and right invariant one-forms are then calculated.  This enables the calculation of the invariant volume element or Haar measure.  These are then used to describe the density matrix of a pure state and geometric phases for three state systems. 
\end{abstract}

\section*{I.  Introduction}

This paper has two main goals:  the first is to construct the invariant vector fields and one-forms on $SU(3)$ in Euler coordinates that are similar to the Euler angle parameters of $SU(3)$; the second is the application of these structures to the description of the geometric phases of three state systems.

The Euler coordinates along with the left and right invariant vector fields and one-forms are constructed here for several reasons.  One, the Euler decomposition of a group element is achieved by a method that is very general and could, in principle, be applied to $SU(n)$.  Two, this particular parameterization is convenient for many calculations as will be shown.  Three, there will be advantages to having a set of invariant vector fields and corresponding one-forms in the same parameterization of the group. 

Color \cite{W}, flavor \cite{efw}, and nuclear models \cite{Elliot} use the group $SU(3)$.  However, geometric aspects of the group manifold have not emphasized.  That is not to say that the differential geometric structures are not relevant to these models; indeed they are!  However, as shown in Refs. \cite{m1}, and \cite{m2}, the differential geometry of $SU(3)$ is an integral part of the description of three state systems.  The Euler coordinates \cite{note1} and differential forms provide a complete description of the differential geometry needed to describe the pure state density matrix and geometric phase \cite{snw} of three state systems.

\section*{II.  The Lie Algebra}

The Lie Algebra of a Lie group is a set of left invariant vector fields on the group manifold.  These can be constructed by a method that has been applied to $SU(2)$ (see for example \cite{lb}).  The method for decomposing a group element into Euler coordinates and the method for constructing the left invariant vector fields can in principle be applied to $SU(n)$.  Here we provide the details for $SU(3)$.

\subsection*{II.1.  Euler Decomposition}

To parameterize the group into the ``Euler angle'' type parameterization, or Euler coordinates, we first consider the Lie algebra of the group.  The algebra obeys the following commutation relations:
$$
\left[ \la_i, \la_j \right] = iC^k_{\;ij} \la_k.
$$
The structure constants, $C^k_{\;ij}$, can be chosen totally antisymmetric in the three indices.  Then the distinction between upper and lower indices is not relevant (One may also use the Killing metric to lower the index).  In the usual, Gell-Mann basis, the non-zero structure constants are given by:
$$
C_{123} = 2, \;\;\;\;\;\;\;\;\; C_{458} = C_{678}= \sqrt{3},
$$
$$
C_{147} = C_{246} = C_{257} = C_{345} = C_{516} = C_{637} = 1.
$$
We can use the Cartan decomposition and split the Lie algebra of $SU(3)$, denoted here $\eL(SU(3))$, into a Lie subalgebra ${\cal L}(K)$ and the set of all other $SU(3)$ Lie algebra elements ${\cal U}(P)$.  The Lie algebra may then be expressed as a (semi-) direct sum,
$$
\eL(G) = \eL(K) \oplus {\cal U}(P).
$$
To this, there corresponds a decomposition of the group,
$$
G = K \cdot P.
$$
Here $\eL(K) = \{ \la_1, \la_2, \la_3, \la_8\}$, and ${\cal U}(P) = \{\la_4, \la_5, \la_6, \la_7 \}$ and according to the Cartan decomposition, they obey, for $k_1, k_2 \in \eL(K)$, and $p_1, p_2 \in {\cal U}(P)$, the following Lie brackets:
$$
\left[ k_1, k_2 \right] \in \eL(K), \;\;\;\;\;\;\;\left[ p_1, p_2 \right] \in \eL(K),
$$
and
$$
\left[ k_1, p_1 \right] \in {\cal U}(P).
$$
Thus $K$ is the $U(2)$ subgroup obtained by exponentiating the corresponding algebra \{$\la_1,\la_2,\la_3$\} plus $\la_8$.  Since $\la_8$ commutes with this $SU(2)$ subgroup, an element of $K$ can be written as
$$
K= e^{(i\la_3 \alpha)} e^{(i\la_2 \beta)}e^{(i\la_3 \gamma)} e^{(i\la_8 \phi)},
$$
by the Euler angle parameterzation of $SU(2)$.

Next, $P$ may be parameterized.  Of the four elements of $P$, we can pick one, say $\la_5$, and write any element of $P$ as 
$$
P = K e^{(i\la_5 \theta)}K,
$$
Dropping the redundancies, we arrive at the following product for $SU(3)$. (This method of was adapted from the book by Robert Hermann \cite{RH} and first pointed out to me by Larry Biedenharn \cite{lb2}.)  
\beq
D(\alpha,\beta,\gamma,\theta,a,b,c,\phi) = e^{(i\la_3 \alpha)} e^{(i\la_2 \beta)}e^{(i\la_3 \gamma)} e^{(i\la_5 \theta)} e^{(i\la_3 a)} e^{(i\la_2 b)} 
e^{(i\la_3 c)} e^{(i\la_8 \phi)},
\eeq
where $D$ is an arbitrary element of $SU(3)$.  This can be written as
$$
D(\alpha,\beta,\gamma,\theta,a,b,c,\phi) = D^{(2)}(\alpha,\beta,\gamma)
e^{(i\la_5 \theta)}D^{(2)}(a,b,c)e^{(i\la_8 \phi)},
$$
where the $D$ denotes an arbitrary element of $SU(3)$ and $D^{(2)}$ is an arbitrary element of $SU(2)$ as a subset of $SU(3)$.  Since $SU(3)$ is semi-simple, it has only one connected component (see for instance \cite{BandD}).  Thus the given parameterization covers the group.

\subsection*{II.2.  A Set of Invariant Vector Fields}

To find the differential operators corresponding to the Lie algebra 
elements, one takes derivatives of $D(\alpha,\beta,\gamma,\theta,a,b,c,\phi)$ in (1) with respect to each of its parameters.  (For brevity, this will be called  $D$, dropping the explicit dependence on the parameters.)  Writing the result of differentiation in the form 
$$
e^A Be^{-A} D.
$$
and using the Campbell-Baker-Hausdorff relation to expand this as
$$
e^A Be^{-A} = B+ [A,B] + \frac{1}{2}[A,[A,B]] + \cdot \cdot \cdot,
$$
the series can be resummed.  Thus linear combinations of the Lie algebra elements with coefficients given by sines and cosines of the parameters are obtained.  The following is an example of such a calculation:
\bea
\frac{\partial D}{\partial \gamma} &=& e^{(i\la_3 \alpha)} e^{(i\la_2 \beta)}i\la_3 e^{(i\la_3 \gamma)} e^{(i\la_5 \theta)} e^{(i\la_3 a)} e^{(i\la_2 b)} 
e^{(i\la_3 c)} e^{(i\la_8 \phi)}\nonumber\\
                                   &=&
e^{(i\la_3 \alpha)}e^{(i\la_2 \beta)}
(i\la_3)e^{(-i\la_2 \beta)}e^{(-i\la_3 \alpha)}D\nonumber\\
&=& i(\la_1 \cos{2\alpha} \sin{2\beta} + \la_2 \sin{2\alpha} \sin{\beta}
 +\la_3 \cos{2\beta})D \nonumber.
\eea
There are eight such derivatives.  These linear combinations may then be expressed as
\bea
\partial_j = b^i_j \la_i,
\eea
where the parameters have been labelled sequentially as they appear in (1), {\it i.e.},  

\bea
\partial_1 &\equiv & \frac{\partial}{\partial \alpha}, \;\;\;\;\;\;\;\;\;\;\;\;\;\;\; 
\partial_2 \equiv \frac{\partial}{\partial \beta}, \;\;\;\;\;\;\;\;\;\;\;\;\;\;\;
\hfil \partial_3 \equiv \frac{\partial}{\partial \gamma}\nonumber \\
\partial_5 &\equiv& \frac{\partial}{\partial a},   \;\;\;\;\;\;\;\;\;\;\;\;\;\;\;
\partial_6 \equiv \frac{\partial}{\partial b}, \;\;\;\;\;\;\;\;\;\;\;\;\;\;\; 
\partial_7 \equiv  \frac{\partial}{\partial c},\;\;\;\;\;\;\;\;\;\;\;\;\;\;\;
\nonumber \\
\partial_4 &\equiv & \frac{\partial}{\partial \theta}, \;\;\;\;\;\;\;\;\;\;\;\;\;\;\;
\partial_8 \equiv \frac{\partial}{\partial \phi}.\;\;\;\;\;\;\;\;\;\;\;\;\;\;\;\nonumber 
\eea

The matrix $ b^i_j$ is then inverted to obtain the representation of the Lie algebra of $SU(3)$ as left invariant vector fields on the group manifold in the Euler parameterization.  Note that this calculation relies only on the commutation relations and not on any particular representation of the algebra.

The left invariant vector fields are differential operators on smooth functions $D^*$ on $SU(3)$; they satisfy \cite{note2}
\beq
 \Lambda_i D^* = -\la_i D^*.
\eeq
Explicit forms of these operators are given in Appendix 1.  The operators acting from the right are found similarly by taking right derivatives.  Note, however, that the right operators obey the commutation relation $[\Lambda^r_i,\Lambda^r_j]=-C^k_{ij}\Lambda^r_k$ since a left translation by a group element is directly related to the right translation of the inverse of this element (see for example \cite{note3}, \cite{cd} or \cite{isham}).  

The calculation of the left invariant vector fields acting on the $D$ matrices was first attempted by T.J. Nelson \cite{N}.  However, no attempt was made to calculate the right invariant vector fields or the corresponding forms.

As stated above, the right differential operators (the differential operators that correspond to this action when acting {\it from} the right) are different and are denoted 
$\Lambda_i^r$.
One may find these ``right'' differential operators in two ways.  First, one may use the relation 
$$
D^*\lambda_i = -\Lambda_i^r D^*,
$$
and do the same calculation as with the left ones.  Second, one may use the fact that
\beq
\Lambda_i^r = R_{i}^{\phantom{*}j}\Lambda_j,
\eeq
where the $R_{i}^{\phantom{*}j}\in SO(8)$ is an element of the adjoint representation of $SU(3)$.  It is therefore a function of the eight parameters above.  The matrix $R$ does not cover $SO(8)$ as is clear from the fact that $SU(3)$ has only eight parameters whereas $SO(8)$ has twenty-eight.

The right invariant vector fields are also given in Appendix A and the matrix $R_{i}^{\phantom{*}j}$ has been calculated explicitly in the Euler coordinates as well(see \cite{mands}).

\section*{III.  Invariant Integration}

\subsection*{III.1.  Invariant Forms}

The left invariant forms for the manifold are dual to the left invariant vector fields.  This duality between the tangent and cotangent vectors may be used to construct left invariant one forms from the left invariant vector fields.  If the left invariant vector fields are written as
$$
\lambda_i = a^j_i \partial_j
$$
and the left invariant forms are expressed as
\bea
\omega^{l} = b^l_k dx^k,
\eea
then
\bea
\delta^l_i = \langle \omega^l,\lambda_i \rangle = b^l_k a^j_i \langle dx^k,\partial_j \rangle = 
b^l_k a^j_i \delta_j^k.
\eea
Therefore, the matrices $b$ and $a$ are inverse transposes of each other.  This is, in fact, the same $b$ given in (2).  In this way, we obtain left (and analogously right) invariant forms on the group manifold.  Thus one may now use these to integrate over the whole space or any subspace of the group manifold by the appropriate wedge product of these forms.  The explicit left and right invariant one-forms are given in Appendix B.

\subsection*{III.2.  Invariant Volume Element}

The group invariant volume element, or Haar measure, may now be computed, using the invariant forms.  To do this, compute the wedge product of the invariant one forms or, equivalently, take the determinant of the matrix $b$ above.  The result is
$$
dV = \sin 2\beta \sin 2b \sin 2\theta \sin^2 \theta\; d\alpha\; d\beta\; d\gamma\;
d\theta\; da\; db\; dc\; d\phi.
$$
This agrees with the result of Holland \cite{hol} but is determined only up to normalization.  The normalization is determined by having $\int dV =1$.  

With the explicit left and right invariant forms one can check that the left invariant volume element is equal, as expected, to the right invariant volume element by direct computation (see for instance \cite{BandD}).

The determination of the ranges of the angles in the Euler coordinates is not a trivial matter.  This is related to determining the group invariant volume which would set the normalization of the volume integral.  The determination of this group invariant volume has been attempted by many authors (see \cite{Marinov} and references therein).  Taking the products of states in the fundamental irreducible representations and enforcing the orthogonality relations among them, one may infer the minimum ranges for which these orthogonality relations hold.  This implies the following ranges of the angles:
$$
0 \leq \alpha,\gamma,a,c < \pi, 
$$
$$
 0 \leq \beta,b,\theta \leq \frac{\pi}{2}, \;\;\;\;\;\;\;\;\;\;\; 0 \leq \phi < \sqrt{3}\pi.
$$
A further discussion of this is given in \cite{mands}.

\section*{IV.  Three State Systems}

Before discussing the geometric phase, some preliminary remarks are in order.  (For a review of geometric phases see \cite{nak}, \cite{snw} or the review article \cite{an}.)  

The general pure state density matrices for three-level systems can be represented in the following way \cite{m1}, \cite{m2}: Let $\psi$ be a state in a three dimensional complex Hilbert space ${\cal H}^{(3)}$.  The density matrix is the matrix $\rho$ described by:
\beq
\rho = \psi \psi^{\dagger} = \frac{1}{3}(\Bid + \sqrt{3}\vec{n}\cdot \vec{\lambda})
\eeq
$$
\psi \in {\cal H}^{(3)}, \;\;\;\;\;\;\;\;\;\;\;\;\;\; (\psi,\psi) = 1.
$$
Here the dagger denotes the hermitian conjugate and $\Bid$ is the $3\times3$ unit matrix.  The dot product of $\vec{n}$, and $\vec{\lambda}$, is the ordinary sum over repeated indices $n^r \lambda_r$.  The eight dimensional unit vector, $\vec{n}$, of real fuctions is restricted to four dimentional subspace as discussed below.  The brackets $(\cdot,\cdot)$ denote the inner product on the space ${\cal H}^{(3)}$.  The $\vec{\lambda}$ represents the eight Gell-Mann matrices.
$$
\begin{array}{crcr}

\lambda_1 = \left( \begin{array}{crcl}
                     0 & 1 & 0 \\
                     1 & 0 & 0 \\
                     0 & 0 & 0   \end{array} \right), &

\lambda_2 = \left( \begin{array}{crcr} 
                     0 & -i & 0 \\
                     i &  0 & 0 \\
                     0 &  0 & 0   \end{array} \right), &

\lambda_3 =  \left( \begin{array}{crcr} 
                     1 &  0 & 0 \\
                     0 & -1 & 0 \\
                     0 &  0 & 0   \end{array} \right), \\

\lambda_4 =  \left( \begin{array}{clcr} 
                     0 & 0 & 1 \\
                     0 & 0 & 0 \\
                     1 & 0 & 0   \end{array} \right), &

\lambda_5 = \left( \begin{array}{crcr} 
                     0 & 0 & -i \\
                     0 & 0 & 0 \\
                     i & 0 & 0   \end{array} \right), &

 \lambda_6 = \left( \begin{array}{crcr} 
                     0 & 0 & 0 \\
                     0 & 0 & 1 \\
                     0 & 1 & 0   \end{array} \right), \\

\lambda_7 = \left( \begin{array}{crcr}
                     0 & 0 & 0 \\
                     0 & 0 & -i \\
                     0 & i & 0   \end{array} \right), &

\lambda_8 = \frac{1}{\sqrt{3}}\left( \begin{array}{crcr} 
                     1 & 0 & 0 \\
                     0 & 1 & 0 \\
                     0 & 0 & -2   \end{array} \right).

\end{array}    
$$
These matrices obey the anticommutation relations
$$
\{\lambda_i,\lambda_j\} = \frac{4}{3}\Bid \delta_{ij} + 2 d_{ijk} \lambda_k.
$$
The nonzero components of the completely symmetric tensor $d_{ijk}$ are
$$
d_{118}= d_{228}= d_{338}= -d_{888}= \frac{1}{\sqrt{3}},\;\;\;\;\; d_{448}= d_{558}= d_{668}= d_{778}=-\frac{1}{2\sqrt{3}},
$$
$$
d_{146}= d_{157}= -d_{247}= d_{256}= d_{344}= d_{355}= -d_{366}= -d_{377}=\half.
$$
Using the $d_{ijk}$, the following symmetric product between two elements in the algebra may be defined:
\beq
(\vec{a} \star \vec{b})_i = \sqrt{3}d_{ijk}a_j b_k.
\eeq
Now the restrictions on the density matrix:
$$
\rho^{\dagger}= \rho = \rho^2  \geq 0, \;\;\;\mbox{and}\;\;  \mbox{Tr}\rho = 1,
$$
are equivalent to the following conditions on $n$:
\beq
\vec{n}^*=\vec{n},  \;\;\;\;\; \vec{n} \cdot \vec{n} =1, \;\;\;\;\; \vec{n} \star \vec{n} = \vec{n}.
\eeq

\subsection*{IV.1.  Pure State Density Matrices}

The density matrix for a pure state, {\it e.g.},
\bea
\rho_0 &=& \frac{1}{3}(\Bid - \sqrt{3}\lambda_8)\nonumber \\
       &=&\left( \begin{array}{crcl}
                     0 & 0 & 0 \\
                     0 & 0 & 0 \\
                     0 & 0 & 1   \end{array} \right),
\eea
can be mapped into an arbitrary pure state density matrix, $\rho$, by an $SU(3)$ transformation:
$$
\rho = D \rho_0 D^{-1},\;\;\;\;\;\;\;\;\;\;\;\;  D \in SU(3).
$$
This mapping can be seen as a projection from $SU(3)$ to $SU(3)/U(2)$ since the upper left $2\times 2$ block of this matrix is invariant under a $U(2)$ subgroup of $SU(3)$.  This subgroup is spanned by the parameters $a,b,c$ and $\phi$.  This is clear from the fact that the subalgebra $K$, defined in section II.1, commutes with $\lambda_8$.  Thus an arbitrary pure state density matrix for a three state system is an element of the coset space $SU(3)/U(2)$.  Moreover, this is expressed in a very natural way in terms of the Euler coordinates of the group manifold coordinates. 

It may also be reexpressed in terms of the adjoint representation of the group, {\it viz.},
$$
\rho = \frac{1}{3}(\Bid - \sqrt{3}D\lambda_8D^{-1}) = \frac{1}{3}(\Bid - \sqrt{3}R_{8}^i\lambda_i)
$$
This enables the identification of the matrix elements $-R_{8}^i$ with the vector $\vec{n}$ in (7) and one can show by direct computation that this vector satisfies the requirements (9).

\subsection*{IV.2.  Geometric Phases for Three State Systems}

Using the Euler coordinates for $SU(3)$, in which
$$
D = e^{(i\la_3 \alpha)} e^{(i\la_2 \beta)}
 e^{(i\la_3 \gamma)} e^{(i\la_5 \theta)} e^{(i\la_3 a)} e^{(i\la_2 b)} 
e^{(i\la_3 c)} e^{(i\la_8 \phi)},
$$
one can write down explicit coordinates for $\psi$.  This is done by first calculating $\rho = |\psi><\psi| = \psi \psi^{\dagger}$ by the method above then, and then use $n_i = \frac{\sqrt{3}}{2}\psi^{\dagger} \lambda_i \psi$ to find $\psi$.  The result is
$$
\psi = e^{-2i\phi/\sqrt{3}} \left( \begin{array}{c}
                       e^{i(\alpha + \gamma)} \cos \beta \sin \theta \\
                       e^{-i(\alpha - \gamma)} \sin \beta \sin \theta \\
                       \cos \theta
                       \end{array} \right).
$$
Using this, one can calculate the connection one-form and curvature two-form.  The connection one-form for the geometric phase is 
\beq
{\cal A} = \psi \;d \psi^{\dagger}.
\eeq 
From this, the connection one-form for the three state system is
\bea   
{\cal A}  &=& -\frac{2}{\sqrt{3}}d\phi + \sin^2 \theta [\cos^2 \beta (d\alpha +d\gamma) - \sin^2 \beta (d\alpha - d\gamma)] \nonumber \\
         &=& -\frac{2}{\sqrt{3}}d\phi + \sin^2 \theta \cos 2\beta \;d\alpha + \sin^2 \theta \;d\gamma,
\eea
and the curvature two-form is given by
\bea
F &=& d{\cal A} = -i d\psi^{\dagger}\wedge d\psi \nonumber \\
  &=& \sin 2\theta \cos^2 \beta d\theta \wedge d(\alpha + \gamma)
      -\sin^2 \theta \sin 2 \beta d\beta \wedge d(\alpha + \gamma) \nonumber \\
  & & -\sin 2 \theta \sin^2 \beta d\theta \wedge d(\alpha - \gamma)
      -\sin^2 \theta \sin 2\beta d\beta \wedge d(\alpha - \gamma) \nonumber \\
  &=&  \sin 2 \theta \cos 2 \beta \;d\theta \wedge d\alpha -2\sin^2 \theta \sin 2 \beta \;d\beta \wedge d\alpha + \sin 2 \theta \;d\theta \wedge d\gamma.
\eea
The geometric phase around a closed loop is given in general by
$$
\varphi_g = \int {\cal A}
$$
and specifically for the three state systems by
\bea
\varphi_g &=&\int \sin^2 \theta [\cos^2 \beta (d\alpha +d\gamma) - \sin^2 \beta (d\alpha - d\gamma)] \nonumber \\
          &=& \int [\sin^2 \theta \cos 2\beta \;d\alpha + \sin^2 \theta \;d\gamma],
\eea
or by the integral of the curvature over the corresponding surface enclosed by this loop.  The pull back from $SU(3)/U(2)$ to $SU(3)$ of the connection one-form, ${\cal A}$, is expressible in terms of linear combinations of the one-forms on the group manifold.  Let $D$ be a point in $SU(3)$ as above, and $\rho$ a point in $SU(3)/U(2)$.  The push forward, (or derivative mapping) and pull back maps denoted $\rho^*$, and $\rho_*$ respectively, give the relationship between the vector fields and one-forms on the group manifold and coset space.

\section*{Conclusions/Comments}

The left and right invariant vector fields on $SU(3)$ in an Euler angle paramterization were displayed explicitly.  This enables the calculation of tensor products by direct computation (see for instance \cite{cd1}).  In the case of three state systems there is a natural projection map from $SU(3)$ to the coset space $SU(3)/U(2)$ that describes the density matrix of pure states.  Having a natural projection map one may use all of the fiber bundle techniques for computing the explicit structures in terms of the Euler coordinates.  This enables the calculation of vector fields and forms on the coset space.  It also provides a means of calculating geometric phases for the three state systems.

It was stated earlier that the methods for calculating the invariant vector fields and forms for $SU(3)$ are generalizable to $SU(n)$.  Though this is true, the left invariant one-forms given here were calculated by hand.  This is obvioulsly not an enviable task for an $SU(n)$ group since the dimension of these groups is $n^2-1$.  It is possible to calculate the invariant vector fields with {\it Mathematica}\cite{dave}.  However, this program has some difficulty simplifying products of trigonometric functions of the form found in this calculation.  As such programs become more proficient at symbolic manipulations of this sort, the calculations and analyses of such structures for higher dimensional Lie groups will become more managable.

\section*{Acknowledgements}

I would, first and foremost, like to thank Prof. L. C. Biedenharn (now deceased) who, as my advisor, first started me on the construction of these structures.  I could not give him too much credit here.  I would also like to thank Prof. E. C. G. Sudarshan for many helpful comments along with Prof. Duane Dicus whose help and support enabled the completion of this paper.  I would like to thank David Cook for deriving the vector fields independently.  This provided a check for the results below.  This research was supported in part by the U.S. Department of Energy under Contract No. DE-EG013-93ER40757.

\section*{Appendix A.}

The left invariant vector fields are as follows:

{\small

\bea
\Lambda_1&=&i \cos2\alpha \cot2\beta \partial_1 + i \sin2\alpha \partial_2 - 
i\frac{\cos2\alpha}{\sin2\beta} \partial_3 \\ 
\Lambda_2&=&-i \sin2\alpha \cot2\beta \partial_1+i\cos2\alpha\partial_2+i
\frac{\sin2\alpha}{\sin2\beta}\partial_3 \\
\Lambda_3&=&i\partial_1 \\
\Lambda_8 &=&i\sqrt{3}\partial_3 - i\sqrt{3}\partial_5 + i\partial_8
\eea
\bea
\Lambda_4&=&i\frac{\sin\beta}{\sin2\beta}\cot\theta \cos(\alpha+\gamma) \partial_1
-i\sin\beta \cot\theta \sin(\alpha+\gamma)\partial_2 \nonumber \\
         &-&i\cot2\beta \sin\beta \cot\theta \cos(\alpha+\gamma)\partial_3
+i\frac{(2-\sin^2\theta)}{\sin2\theta}\cos\beta \cos(\alpha+\gamma)\partial_3\nonumber \\
	 &+&i\cos\beta \sin(\alpha+\gamma)\partial_4\nonumber \\
         &-&i2\frac{\cos\beta}{\sin2\theta} \cos(\alpha + \gamma)\partial_5
-i\frac{\cot2b}{\sin\theta}\sin\beta \cos(\alpha-\gamma-2a)\partial_5\nonumber \\
         &+&i\frac{\sin\beta}{\sin\theta}\sin(\alpha-\gamma-2a)\partial_6\nonumber \\
         &+&i\frac{\sin\beta}{\sin\theta \sin2b}\cos(\alpha-\gamma-2a)\partial_7\nonumber \\
         &-&\frac{\sqrt{3}}{2}\tan\theta \cos\beta \cos(\alpha+\gamma)\Lambda_8
\eea
\bea
\Lambda_5&=&-i\frac{\sin\beta}{\sin2\beta}\cot\theta \sin(\alpha+\gamma) \partial_1
            -i\sin\beta \cot\theta \cos(\alpha+\gamma)\partial_2 \nonumber \\
         &+&i\cot2\beta \sin\beta \cot\theta \sin(\alpha+\gamma)\partial_3
          -i\frac{(2-\sin^2\theta)}{\sin2\theta}\cos\beta \sin(\alpha+\gamma)\partial_3\nonumber\\
         &+&i\cos\beta \cos(\alpha+\gamma)\partial_4\nonumber \\
         &+&i2\frac{\cos\beta}{\sin2\theta} \sin(\alpha + \gamma)\partial_5
          +i\frac{\cot2b}{\sin\theta}\sin\beta \sin(\alpha-\gamma-2a)\partial_5\nonumber \\
         &+&i\frac{\sin\beta}{\sin\theta}\cos(\alpha-\gamma-2a)\partial_6\nonumber \\
         &-&i\frac{\sin\beta}{\sin\theta \sin2b}\sin(\alpha-\gamma-2a)\partial_7\nonumber \\
         &+&\frac{\sqrt{3}}{2}\tan\theta \cos\beta \sin(\alpha+\gamma)\Lambda_8
\eea
\bea
\Lambda_6&=&i\frac{\cos\beta}{\sin2\beta}\cot\theta \cos(\alpha-\gamma) \partial_1
            +i\cos\beta \cot\theta \sin(\alpha-\gamma)\partial_2 \nonumber \\
         &-&i\cot2\beta \cos\beta \cot\theta \cos(\alpha-\gamma)\partial_3
          -i\frac{(2-\sin^2\theta)}{\sin2\theta}\sin\beta \cos(\alpha-\gamma)\partial_3\nonumber \\
	 &+&i\sin\beta \sin(\alpha-\gamma)\partial_4\nonumber \\
         &+&i2\frac{\sin\beta}{\sin2\theta} \cos(\alpha - \gamma)\partial_5
          -i\frac{\cot2b}{\sin\theta}\cos\beta \cos(\alpha+\gamma+2a)\partial_5\nonumber \\
         &-&i\frac{\cos\beta}{\sin\theta}\sin(\alpha+\gamma+2a)\partial_6\nonumber \\
         &+&i\frac{\cos\beta}{\sin\theta \sin2b}\cos(\alpha+\gamma+2a)\partial_7\nonumber \\
         &+&\frac{\sqrt{3}}{2}\tan\theta \sin\beta \cos(\alpha-\gamma)\Lambda_8
\eea
\bea
\Lambda_7&=&i\frac{\cos\beta}{\sin2\beta}\cot\theta \sin(\alpha-\gamma) \partial_1
           -i\cos\beta \cot\theta \cos(\alpha-\gamma)\partial_2 \nonumber \\
         &-&i\cot2\beta \cos\beta \cot\theta \sin(\alpha-\gamma)\partial_3
          -i\frac{(2-\sin^2\theta)}{\sin2\theta}\sin\beta \sin(\alpha-\gamma)\partial_3\nonumber \\
         &-&i\sin\beta \cos(\alpha-\gamma)\partial_4\nonumber \\
         &+&i2\frac{\sin\beta}{\sin2\theta} \sin(\alpha - \gamma)\partial_5
          -i\frac{\cot2b}{\sin\theta}\cos\beta \sin(\alpha+\gamma+2a)\partial_5\nonumber \\
         &+&i\frac{\cos\beta}{\sin\theta}\cos(\alpha+\gamma+2a)\partial_6\nonumber \\
         &+&i\frac{\cos\beta}{\sin\theta \sin2b}\sin(\alpha+\gamma+2a)\partial_7\nonumber \\
         &+&\frac{\sqrt{3}}{2}\tan\theta \sin\beta \sin(\alpha-\gamma)\Lambda_8
\eea

}

Where $\eta = \phi/ \sqrt{3}$. 
  The right invariant vector fields are given by the following equations.

{\small

\bea
\Lambda^r_1&=&-i\cos 2c \cot 2b \partial_7 -i\sin 2c \partial_6 +i 
\frac{\cos 2c}{\sin 2b}\partial_5 \\
\Lambda_2^r&=&-i\sin 2c \cot 2b \partial_7 + i\cos 2c \partial_6+i
\frac{\sin 2c}{\sin 2b}\partial_5 \\
\Lambda_3^r&=&i\partial_7\\
\Lambda_8^r&=&i\partial_8
\eea
\bea
\Lambda_4^r&=&-i\frac{\sin b}{\sin 2b}\cot \theta \cos(c+a+3\eta) \partial_7\nonumber\\
           &+&i\sin b \cot \theta \sin(c+a+3\eta) \partial_6\nonumber\\
           &+&i\cot 2b \sin b \cot \theta \cos(c+a+3\eta) \partial_5
            -i\frac{(2-\sin^2 \theta)}{\sin 2 \theta} \cos b \cos(c+a+3\eta)\partial_5\nonumber\\
           &-&i\cos b \sin(c+a+3\eta) \partial_4\nonumber\\
           &+&i2\frac{\cos b}{\sin 2\theta}\cos(c+a+3\eta)\partial_3
            +i\frac{\cot 2 \beta}{\sin \theta}\sin b\cos(c-a-2\gamma+3\eta)\partial_3\nonumber\\
           &-&i\frac{\sin b}{\sin \theta}\sin(c-a-2\gamma+3\eta)\partial_2\nonumber\\
           &-&i\frac{\sin b}{\sin \theta \sin 2\beta} \cos(c-a-2\gamma +3\eta)
\partial_1\nonumber\\
           &-&\frac{\sqrt{3}}{2}\tan \theta \cos b \cos(c+a+3\eta)\Lambda_8^r
\eea
\bea
\Lambda_5^r&=&-i\frac{\sin b}{\sin 2b}\cot \theta \sin(c+a+3\eta) \partial_7\nonumber\\
           &-&i\sin b \cot \theta \cos(c+a+3\eta) \partial_6\nonumber\\
           &+&i\cot 2b \sin b \cot \theta \sin(c+a+3\eta) \partial_5
            -i\frac{(2-\sin^2 \theta)}{\sin 2 \theta} \cos b \sin(c+a+3\eta)\partial_5\nonumber\\
           &+&i\cos b \cos(c+a+3\eta) \partial_4\nonumber\\
           &+&i2\frac{\cos b}{\sin 2\theta}\sin(c+a+3\eta)\partial_3
            +i\frac{\cot 2 \beta}{\sin \theta}\sin b\sin(c-a-2\gamma+3\eta)\partial_3\nonumber\\
           &+&i\frac{\sin b}{\sin \theta}\cos(c-a-2\gamma+3\eta)\partial_2\nonumber\\
           &-&i\frac{\sin b}{\sin \theta \sin 2\beta} \sin(c-a-2\gamma +3\eta)
\partial_1\nonumber\\
           &-&\frac{\sqrt{3}}{2}\tan \theta \cos b \sin(c+a+3\eta)\Lambda_8^r
\eea

\bea
\Lambda_6^r&=&i\frac{\cos b}{\sin 2b}\cot \theta \cos(c-a-3\eta) \partial_7\nonumber\\
           &+&i\cos b \cot \theta \sin(c-a-3\eta) \partial_6\nonumber\\
           &-&i\cot 2b \cos b \cot \theta \cos(c-a-3\eta) \partial_5
            -\frac{(2-\sin^2 \theta)}{\sin 2 \theta} \sin b \cos(c-a-3\eta)\partial_5\nonumber\\
           &+&i\sin b \sin(c-a-3\eta) \partial_4\nonumber\\
           &+&i2\frac{\sin b}{\sin 2\theta}\cos(c-a-3\eta)\partial_3
            -i\frac{\cot 2 \beta}{\sin \theta}\cos b\cos(c+a+2\gamma-3\eta)\partial_3\nonumber\\
           &-&i\frac{\cos b}{\sin \theta}\sin(c+a+2\gamma-3\eta)\partial_2\nonumber\\
           &+&i\frac{\cos b}{\sin \theta \sin 2\beta} \cos(c+a+2\gamma -3\eta)
\partial_1\nonumber\\
           &-&\frac{\sqrt{3}}{2}\tan \theta \sin b \cos(c-a-3\eta)\Lambda_8^r
\eea

\bea
\Lambda_7^r&=&-i\frac{\cos b}{\sin 2b}\cot \theta \sin(c-a-3\eta) \partial_7\nonumber\\
           &+&i\cos b \cot \theta \cos(c-a-3\eta) \partial_6\nonumber\\
           &+&i\cot 2b \cos b \cot \theta \sin(c-a-3\eta) \partial_5
            +i\frac{(2-\sin^2 \theta)}{\sin 2 \theta} \sin b \sin(c-a-3\eta)\partial_5\nonumber\\
           &+&i\sin b \cos(c-a-3\eta) \partial_4\nonumber\\
           &-&i2\frac{\sin b}{\sin 2\theta}\sin(c-a-3\eta)\partial_3
            +i\frac{\cot 2 \beta}{\sin \theta}\cos b\sin(c+a+2\gamma-3\eta)\partial_3\nonumber\\
           &-&i\frac{\cos b}{\sin \theta}\cos(c+a+2\gamma-3\eta)\partial_2\nonumber\\
           &-&i\frac{\cos b}{\sin \theta \sin 2\beta} \sin(c+a+2\gamma -3\eta)
\partial_1\nonumber\\
           &+&\frac{\sqrt{3}}{2}\tan \theta \sin b \sin(c-a-3\eta)\Lambda_8^r 
\eea

}

\section*{Appendix B.  Invariant Forms}

The invariant forms on the manifold are given by the following equations.

{\small

\bea
\omega^1&=&
-i\sin (2\,\alpha )d\beta +i 
  \cos (2\,\alpha )\,\sin (2\,\beta )d\gamma \nonumber \\
&+& 
  i\cos (2\,\alpha )\,\sin (2\,\beta )\,
    \left( 1 - \half{\sin^2 (\theta) } \right) da   \nonumber \\
&-& i\cos (2\,a + 2\,\gamma )\,\cos (\theta )\,
      \sin (2\,\alpha ) db \nonumber \\
&-& i\cos (2\,\alpha )\,\cos (2\,\beta )\,\cos (\theta )\,
      \sin (2\,a + 2\,\gamma ) db   \nonumber \\
&+& i\left[ \cos (2\,\alpha )\,
        \cos (2\,\beta )\,\cos (2\,a + 2\,\gamma )\,
        \cos (\theta )\,\sin (2\,b) \right]dc \nonumber \\
&-& i\cos (\theta )\,\sin (2\,\alpha )\,\sin (2\,b)\,
      \sin (2\,a + 2\,\gamma ) dc \nonumber \\
&+& i\cos (2\,\alpha )\,\cos (2\,b)\,\sin (2\,\beta )\,
      \left( 1 - \half{\sin^2 (\theta) } \right) dc \nonumber \\
&-& i\frac{\sqrt{3}}{2}\cos (2\,\alpha )\,
      \sin (2\,\beta )\,\sin^2 \theta d\phi \\
\omega^2&=&
-i\cos (2\,\alpha )d\beta - 
  i\sin (2\,\alpha )\,\sin (2\,\beta )d\gamma \nonumber \\
&-&  i\sin (2\,\alpha )\,\sin (2\,\beta )\,
   \left( 1 - \half\sin^2 (\theta ) \right)da  \nonumber \\
&-&  i\cos (2\,\alpha )\,
      \cos (2\,a + 2\,\gamma )\,\cos (\theta ) db \nonumber \\
&+&  i\cos (2\,\beta )\,\cos (\theta )\,\sin (2\,\alpha )\,
      \sin (2\,a + 2\,\gamma ) db  \nonumber \\
&-&  i\cos (2\,\beta )\,\cos (2\,a + 2\,\gamma )\,
      \cos (\theta )\,\sin (2\,\alpha )\,\sin (2\,b) dc \nonumber \\
&-&  i\cos (2\,\alpha )\,\cos (\theta )\,\sin (2\,b)\,
      \sin (2\,a + 2\,\gamma ) dc \nonumber \\
&-&  i\cos (2\,b)\,\sin (2\,\alpha )\,\sin (2\,\beta )\,
      \left( 1 - \half\sin^2 (\theta ) \right) dc  \nonumber \\
&+&  i\frac{\sqrt{3}}{2}\sin (2\,\alpha )\,\sin (2\,\beta )\,
      \sin^2 (\theta )d\phi \\
\omega^3&=&
-id\alpha - i\cos (2\,\beta )d\gamma  \nonumber \\
&-&  i\cos (2\,\beta )\,
   \left( 1 - \half\sin^2 (\theta ) \right)da   \nonumber \\
&-&  i\cos (\theta )\,\sin (2\,\beta )\,
   \sin (2\,a + 2\,\gamma )db  \nonumber \\
&+& i\cos (2\,a + 2\,\gamma )\,
        \cos (\theta )\,\sin (2\,b)\,\sin (2\,\beta ) dc \nonumber \\
&-& i\cos (2\,b)\,\cos (2\,\beta )\,
      \left( 1 - \half\sin^2 (\theta ) \right) dc \nonumber \\
&+& i\frac{\sqrt{3}}{2}\cos (2\,\beta )\,
      \half\sin^2 (\theta )d\phi \\
\omega^4&=&
   -i\cos (\beta )\,\sin (\alpha  + \gamma )d\theta \nonumber \\
&+&  i\half\cos (\beta )\,
      \cos (\alpha  + \gamma )\,\sin (2\,\theta )da  \nonumber \\
&+&  i\sin (\beta )\,\sin (2\,a - \alpha  + \gamma )\,
   \sin (\theta )db  \nonumber \\
&-&  i\cos (2\,a - \alpha  + \gamma )\,
      \sin (2\,b)\,\sin (\beta )\,\sin (\theta ) dc \nonumber \\
&+&  i\half\cos (2\,b)\,\cos (\beta )\,\cos (\alpha  + \gamma )\,
         \sin (2\,\theta ) dc \nonumber \\
&+&  i\frac{\sqrt{3}}{2}\cos (\beta )\,
      \cos (\alpha  + \gamma )\,\sin (2\,\theta )d\phi \\
\omega^5&=&
-i\cos (\beta )\,\cos (\alpha  + \gamma )d\theta  \nonumber \\
&-& i\half\cos (\beta )\,
      \sin (\alpha  + \gamma )\,\sin (2\,\theta )da  \nonumber \\
&-& 
  i\cos (2\,a - \alpha  + \gamma )\,\sin (\beta )\,
   \sin (\theta )db  \nonumber \\
&-& i\sin (2\,b)\,\sin (\beta )\,
      \sin (2\,a - \alpha  + \gamma )\,\sin (\theta ) dc \nonumber \\
&-& i\half\cos (2\,b)\,\cos (\beta )\,\sin (\alpha  + \gamma )\,
         \sin (2\,\theta ) dc \nonumber\\
&-& i\frac{\sqrt{3}}{2}\cos (\beta )\,
      \sin (\alpha  + \gamma )\,\sin (2\,\theta )d\phi \\
\omega^6&=&
-i\sin (\beta )\,\sin (\alpha  - \gamma )d\theta  \nonumber \\
&+& 
  i\cos (\beta )\,\sin (2\,a + \alpha  + \gamma )\,
   \sin (\theta )db \nonumber \\
&-& i\half\cos (\alpha  - \gamma )\,
      \sin (\beta )\,\sin (2\,\theta )da  \nonumber \\
&-& 
  i\frac{\sqrt{3}}{2}\cos (\alpha  - \gamma )\,
      \sin (\beta )\,\sin (2\,\theta )d\phi  \nonumber \\
&-&  i\cos (\beta )\,
      \cos (2\,a + \alpha  + \gamma )\,\sin (2\,b)\,
      \sin (\theta ) dc \nonumber \\
&-& i\half\cos (2\,b)\,
         \cos (\alpha  - \gamma )\,\sin (\beta )\,
         \sin (2\,\theta ) dc \\
\omega^7&=&
+i\cos (\alpha  - \gamma )\,\sin (\beta ) d\theta   \nonumber \\
&-&  i\half\sin (\beta )\,\sin (\alpha  - \gamma )\,
      \sin (2\,\theta )da  \nonumber \\
&-&  i\cos (\beta )\,
   \cos (2\,a + \alpha  + \gamma )\,\sin (\theta )db \nonumber \\
&-&  i\cos (\beta )\,\sin (2\,b)\,
      \sin (2\,a + \alpha  + \gamma )\,\sin (\theta ) dc \nonumber \\
&-&  i\half\cos (2\,b)\,\sin (\beta )\,\sin (\alpha  - \gamma )\,
         \sin (2\,\theta ) dc \nonumber \\
&-&  i\frac{\sqrt{3}}{2}\sin (\beta )\,
      \sin (\alpha  - \gamma )\,\sin (2\,\theta ) d\phi \\
\omega^8 &=&
i\frac{\sqrt{3}}{2}\sin^2 (\theta )da
+ i\frac{\sqrt{3}}{2}\cos (2\,b)\,
      \sin^2 (\theta )dc
- i\left( 1 - \frac{3}{2}\sin^2 (\theta )
      \right)d\phi
\eea

}

The right invariant forms are given by:
{\small

\bea
\omega^1_r&=&
     i\cos (2\,b)\,\cos (2\,c)\,
        \cos (2\,a + 2\,\gamma )\,\cos (\theta )\,
        \sin (2\,\beta )d\alpha    \nonumber \\
&-&  i\cos (\theta )\,\sin (2\,\beta )\,\sin (2\,c)\,
      \sin (2\,a + 2\,\gamma ) d\alpha \nonumber \\
&+&  i\cos (2\,\beta )\,\cos (2\,c)\,\sin (2\,b)\,
      \left( 1 - \half\sin^2 (\theta ) \right) d\alpha  \nonumber \\
&+&  i\cos (2\,c)\,\sin (2\,b)\,
   \left( 1 - \half\sin^2 (\theta ) \right)d\beta   \nonumber \\
&-&  i\cos (2\,a + 2\,\gamma )\,\cos (\theta )\,
      \sin (2\,c) d\gamma \nonumber \\ 
&-&  i\cos (2\,b)\,\cos (2\,c)\,\cos (\theta )\,
      \sin (2\,a + 2\,\gamma ) d\gamma    \nonumber \\
&+&  i\cos (2\,c)\,\sin (2\,b)\,
   \left( 1 - \half\sin^2 (\theta ) \right) d\theta 
 - i\sin (2\,c)db \\
\omega^2_r&=&
  -i\cos (2\,b)\,\cos (2\,a + 2\,\gamma )\,
      \cos (\theta )\,\sin (2\,\beta )\,\sin (2\,c) d\alpha \nonumber \\
&-& 
     i\cos (2\,c)\,\cos (\theta )\,\sin (2\,\beta )\,
      \sin (2\,a + 2\,\gamma ) d\alpha   \nonumber \\
&-&  i\cos (2\,\beta )\,\sin (2\,b)\,\sin (2\,c)\,
      \left( 1 - \half\sin^2 (\theta ) \right) d\alpha  \nonumber \\
&-&  i\sin (2\,b)\,\sin (2\,c)\,
   \left( 1 - \half\sin^2 (\theta ) \right)d\beta    \nonumber \\
&-&  i\left[ \cos (2\,c)\,\cos (2\,a + 2\,\gamma )\,
      \cos (\theta ) - \cos (2\,b)\,\cos (\theta )\,
      \sin (2\,c)\,\sin (2\,a + 2\,\gamma ) \right]d\gamma    \nonumber \\
&-&  i\sin (2\,b)\,\sin (2\,c)\,
   \left( 1 - \half\sin^2 (\theta ) \right)d\theta + \cos (2\,c)db \\
\omega^3_r&=&  i\cos (2\,a + 2\,\gamma )\,
    \cos (\theta )\,\sin (2\,b)\,\sin (2\,\beta ) d\alpha \nonumber \\
&-& i\cos (2\,b)\,\cos (2\,\beta )\,
      \left( 1 - \half\sin^2 (\theta ) \right) d\alpha  \nonumber \\
&-& i\cos (2\,b)\,
    \left( 1 - \half\sin^2 (\theta ) \right) d\beta \nonumber \\ 
&-& i\cos (\theta )\,\sin (2\,b)\,
   \sin (2\,a + 2\,\gamma )d\gamma   \nonumber \\
&-& i\cos (2\,b)\,
   \left( 1 - \half\sin^2 (\theta ) \right)d\theta + dc  \\
\omega^4_r&=& -i\cos (a - c + 2\,\gamma  -3\eta )\,\sin (b)\,
      \sin (2\,\beta )\,\sin (\theta ) d\alpha \nonumber \\ 
&+& i\half\cos (b)\,\cos (2\,\beta )\,\cos (a + c +3\eta )\,
         \sin (2\,\theta )d\alpha    \nonumber \\
&+& i\half\cos (b)\,\cos (a + c +3\eta )\,
        \sin (2\,\theta ) d\beta \nonumber \\ 
&+& i\sin (b)\,\sin (\theta )\,
   \sin (a - c + 2\,\gamma  - 3\eta)d\gamma  \nonumber \\
&+& i\half\cos (b)\,\cos (a + c + 3\eta)\,
      \sin (2\,\theta )d\theta 
- \cos (b)\,\sin (a + c + 3\eta)da \\
\omega^5_r&=& -i\sin (b)\,\sin (2\,\beta )\,
      \sin (\theta )\,\sin (a - c + 2\,\gamma  - 3\eta) d\alpha \nonumber \\
&-& i\half\cos (b)\,\cos (2\,\beta )\,\sin (2\,\theta )\,
         \sin (a + c +3\eta ) d\alpha  \nonumber \\
&-& i\half\cos (b)\,\sin (2\,\theta )\,
      \sin (a + c + 3\eta) d\beta \nonumber \\
&-& i\cos (a - c + 2\,\gamma  - 3\eta)\,\sin (b)\,
   \sin (\theta )d\gamma   \nonumber \\
&-& i\half\cos (b)\,\sin (2\,\theta )\,
      \sin (a + c + 3\eta)d\theta 
-i\cos(b)\,\cos(a+c+3\eta)da \\
\omega^6_r&=&  -i\cos (b)\,\cos (a + c + 2\,\gamma  - 3\eta)\,
      \sin (2\,\beta )\,\sin (\theta ) d\alpha \nonumber \\
&-& i\half\cos (2\,\beta )\,\cos (a - c + 3\eta)\,\sin (b)\,
         \sin (2\,\theta ) d\alpha    \nonumber \\
&-& i\half\cos (a - c +3\eta )\,\sin (b)\,
      \sin (2\,\theta )d\beta \nonumber \\
&-& i\cos (b)\,\sin (\theta )\,
   \sin (a + c + 2\,\gamma  - 3\eta)d\gamma   \nonumber \\
&-& i\half\cos (a - c + 3\eta)\,\sin (b)\,
      \sin (2\,\theta )d\theta 
 + i\sin (b)\,\sin (a - c +3\eta )da \\
\omega^7_r&=& -i\cos (b)\,\sin (2\,\beta )\,
      \sin (\theta )\,\sin (a + c + 2\,\gamma  - 3\eta) d\alpha \nonumber \\
&+&  i\half\cos (2\,\beta )\,\sin (b)\,\sin (2\,\theta )\,
         \sin (a - c + 3\eta)  d\alpha  \nonumber \\
&+&  i\half\sin (b)\,\sin (2\,\theta )\,
      \sin (a - c +3\eta )d\beta  \nonumber \\
&+&  i\cos (b)\,\cos (a + c + 2\,\gamma  - 3\eta)\,
   \sin (\theta )d\gamma  \nonumber \\
&+&  i\half\sin (b)\,\sin (2\,\theta )\,
      \sin (a - c +3\eta )d\theta 
 + i\cos (a - c + 3\eta)\,\sin (b) da \\
\omega^8_r&=&
   i\frac{\sqrt{3}}{2}\cos (2\,\beta )\,
      \sin^2 (\theta )d\alpha
 + i\frac{\sqrt{3}}{2}\sin^2 (\theta ) d\beta
 + i\frac{\sqrt{3}}{2}\sin^2 (\theta )d\theta 
 - id\phi
\eea

}

\end{document}